\documentclass{elsart}
\usepackage{natbib}
\begin{document}
\runauthor{Simpson}
\begin{frontmatter}
\title{Quasars are more luminous than radio galaxies -- so what?}
\author{Chris Simpson}
\address{Subaru Telescope}
\begin{abstract}
Surveys to find high-redshift radio galaxies deliberately exclude
optically-bright objects, which may be distant radio-loud quasars. In order
to properly determine the space density of supermassive black holes, the
fraction of such objects missed must be determined within a quantitative
framework for AGN unification. I briefly describe the receding torus model,
which predicts that quasars should have more luminous ionizing continua
than radio galaxies of similar radio luminosity, and present evidence to
support it. I also suggest two further tests of the model which should
constrain some of its parameters.
\end{abstract}
\begin{keyword}
galaxies: active -- radio continuum: galaxies
\end{keyword}
\end{frontmatter}

\section{Introduction}

As the cosmological applications of radio galaxies become clearer, it is
important to realize that they do not represent a fundamental class of
object. Searches for high-redshift radio sources use spectral index and
optical magnitude selection criteria which result in the exclusion of
radio-loud quasars even though these are fundamentally the same objects,
according to the standard AGN unification paradigm. While this does not
matter if one is using radio galaxies simply as signposts to locate and
study places where large-scale structure is developing
\cite{TNJ1338,SR2002}, it is possible that they could be used to measure
the rate of formation of (spinning) supermassive black holes and/or
clusters. To do this, however, requires not just an understanding of radio
source physics \cite{BRW}, but also a renormalization to account for the
missed quasar population. A \textit{quantitative\/} understanding of how
orientation affects the observed properties of extragalactic radio sources
is therefore required to fully exploit these objects.

\section{The receding torus}

The ubiquity of the `big red bump' longward of $1\,\mu$m in the spectra of
QSOs \cite{QED}, together with the interpretation of this as thermal
emission from hot dust \cite{Barvainis,Fairall9}, seems to indicate that
dust will always exist as close to the active nucleus as physics allows.
If this is the same dust that is responsible for hiding the nucleus from
direct view in narrow-line objects, then it must lie further from the
nucleus in objects with higher ionizing luminosities. The assumption that
the height of the obscuring structure (`torus' in unification parlance)
remains constant leads to the receding torus model \cite{L91,HGD,Bart}. In
this scenario, more luminous objects have a higher probability of being
observed as quasars, and consequently the mean ionizing luminosity of
quasars in an orientationally-unbiased (radio-selected) sample will be
higher than that of radio galaxies. It therefore follows that any quantity
which is more strongly correlated with ionizing luminosity than with radio
luminosity will also be higher in quasars than radio galaxies
\cite{Bart}. Indeed, studies of [O{\sc~iii}] \cite{JB90,L94},
mid-to-far-infrared \cite{MFIR,vanB}, and submillimetre \cite{CJW850}
luminosities indicate that quasars are indeed more luminous than radio
galaxies in these properties by factors of a few. Since the receding torus
model was first described at around the same time as the earliest of these
studies, it has always surprised me that it did not gain greater acceptance
as a way to explain these differences, which were incompatible with a
picture where the torus opening angle was the same in all objects. The lack
of a difference in [O{\sc~ii}] luminosities between quasars and radio
galaxies \cite{Hes} is due to the insensitivity of this line to the
strength of the ionizing continuum \cite{Bart}, while the apparent lack of
a difference in [O{\sc~iii}] luminosity at high redshift \cite{JR97} is
probably a result of large measurement uncertainties --- the lack of a
significant difference does not equate to the two classes having the same
luminosity, and the data are also consistent with quasars being twice as
luminous in this line, as is seen at lower redshift \cite{L94}.

\section{Observational evidence}

Of course, the key assumption in the receding torus model is that the
height of the torus is independent of AGN luminosity. While this seems to
be a reasonable zeroth-order assumption, it need not be true since the AGN
luminosity is correlated with black hole mass, and therefore also with the
mass of the host galaxy; it is quite possible that either or both of these
quantities might affect the height of the torus. However, several pieces of
observational evidence indicate that the height of the torus is not a
strong function of luminosity.

\begin{figure}
\vspace{240pt}
\includegraphics{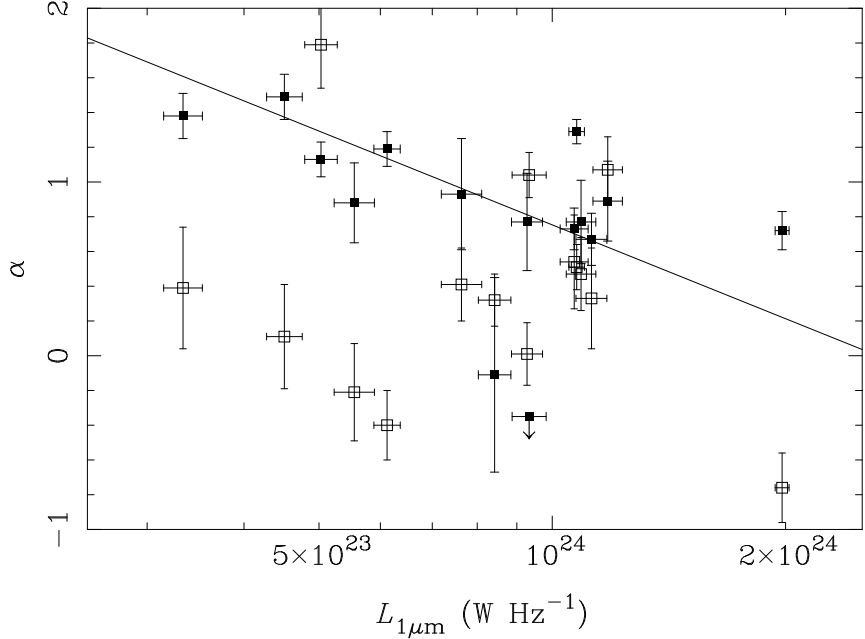}
\caption{Relationship between optical and near-IR spectral indices (open
and solid points, respectively) and rest-frame $1\,\mu$m luminosity for a
complete sample of 3CRR quasars \cite{SR2000}. The solid line indicates the
relationship predicted by the receding torus model ($L_{\rm2\mu m} \propto
L_{\rm1\mu m}^{0.5}$) for the near-IR spectral index \textit{only\/} (with
arbitrary vertical normalization).}
\label{fig:alpha}
\end{figure}

\begin{enumerate}
\item The quasar fraction in different samples should increase with
AGN luminosity as the inner wall of the torus is pushed away
\cite{QSOfrac,Jenny}.
\item The fraction of lightly-reddened quasars (i.e., those where broad
wings are seen on H$\alpha$ but not H$\beta$) should decrease with
luminosity as the solid angle over which lines of sight `graze' the edge of
the torus decreases \cite{SRL99}.
\item The strength of `big red bump' relative to the ionizing continuum
should be less in more luminous objects as the solid angle subtended by the
torus decreases \cite{SR2000,DLG}. This is the cleanest and most
quantitative test and the model fits the data well
(Fig.~\ref{fig:alpha}).
\end{enumerate}

Interestingly, there is also evidence that in the most luminous quasars,
where the inner wall of the torus is pushed beyond a few parsecs, that the
torus disintegrates completely \cite{DLG}. The most distant SDSS quasars
therefore provide an accurate census of the number of supermassive black
holes in the early Universe.

\section{Implications}

None of the pieces of evidence presented in the previous section is
consistent with what some authors refer to as the `simplest' unification
scenario, where broad- and narrow-line objects are separated by a single
critical angle, independent of AGN properties. This should not be a cause
of concern, since such a scenario is unrealistic, requiring either the
inner walls of the torus to unphysically remain at the same distance from
the nucleus irrespective of the AGN luminosity, or the torus height to
increase in a contrived manner so as to maintain a constant opening
angle. In addition, the first and third points above favour a torus rather
than a warped disk since together they indicate that the obscuration is
caused by the hot dust close to the nucleus, rather than by a warp at
larger distances; the strong correlation between nuclear extinction and
viewing angle in narrow-line radio galaxies also favours a torus rather
than a warped disk \cite{SWW2000}.

As explained earlier, the receding torus model results in the mean
(ionizing) luminosity of quasars in a sample being brighter than the mean
(ionizing) luminosity of radio galaxies. The factor by which quasars are
more luminous depends on two quantities: (i) the mean opening angle of the
torus in the sample, which is fixed by the observed quasar fraction, and
(ii) the spread in the distribution of ionizing luminosities in the
sample. This second quantity is impossible to measure, but can be estimated
as the convolution of the observed radio luminosity distribution with a
Gaussian representing the scatter in the radio--ionizing luminosity
correlation. Unfortunately, this scatter is also unknown; I previously
adopted 0.6\,dex \cite{Bart} from the scatter in the radio--optical
correlation \cite{SBGS}, although the use of $M_B$ as a tracer of the total
ionizing luminosity adds uncertainties due to differences in rest-frame
wavelength, extinction, and optical/UV spectral index; tighter
correlations do exist \cite{RS91} and I believe the true scatter is lower.
For now, I shall leave this as a free parameter.

\begin{figure}
\vspace{240pt}
\includegraphics{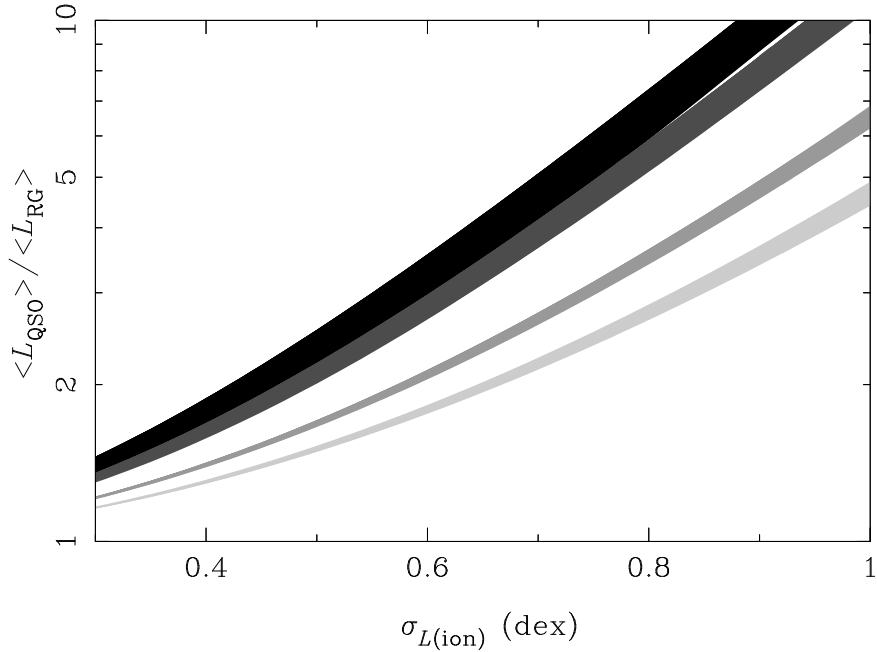}
\caption{Factor by which quasars are more luminous than radio galaxies in
an orientation-independent sample, as a function of the dispersion (assumed
to be a Gaussian in log space) in the ionizing luminosities. The solid
black region is for the simple receding torus model, where the spread
represents samples with different quasar fractions from 30--80\%. The other
regions indicate what happens if there is a random spread in the heights of
the tori in the sample, of factors of 2, 5, and 10 ($1\sigma$ scatter),
with 10 being the lightest gray colour.}
\label{fig:sowhat}
\end{figure}

The solid black region in Fig.~\ref{fig:sowhat} indicates that quasars
exceed radio galaxies by a factor of 2--20 for reasonable values of the
$L_{\rm rad}$--$L_{\rm ion}$ dispersion (also, this number is insensitive
to the quasar fraction)\footnote{Note that fig.~4 in ref.~\cite{Bart} is
incorrect due to improper normalization.}. These values are larger than the
factors of 2--5 by which quasars are observed to be more luminous than
radio galaxies \cite{MFIR,L94,vanB,CJW850}, but this is due to the
simplistic assumption that all tori have the same height. If torus heights
are drawn from a log-normal distribution, the overluminosity is reduced
since it increases the likelihood of low-$L_{\rm ion}$ objects being seen
as quasars (since some will have short tori) while decreasing the
likelihood of high-$L_{\rm ion}$ objects being seen as quasars (since some
will have tall tori). This is represented by the gray regions in
Fig.~\ref{fig:sowhat} where it can be seen that a significant difference in
luminosities persists even when the torus height is allowed to vary by
1\,dex. It is therefore inevitable that quasars will be, on average, more
luminous in their ionizing radiation than radio galaxies, and hence more
luminous in any related quantities (e.g., emission line and infrared
luminosities, which arise from the reprocessing of ionizing photons).  This
is exactly what is observed, and adds additional support for the receding
torus.

\section{Further studies}

None of the observational results quoted in this paper arose from a
deliberate attempt to test the receding torus model. However, the success
of this model, together with the intriguing possibility that
Fig.~\ref{fig:sowhat} could be used to estimate the scatter in the $L_{\rm
rad}$--$L_{\rm ion}$ correlation and the range of torus heights, makes
further study worthwhile.

First, the correlation in Fig.~\ref{fig:alpha} should be extended to lower
luminosities by studying fainter samples of radio-loud quasars at the same
redshift (e.g., from the 7C sample); a sufficiently large number can be
studied with an 8-m telescope in a few nights.

Second, direct evidence for an increase in the opening angle of the torus
should be sought by looking at the distributions of the core-to-lobe ratio,
$R$, in different samples. This measurement can be converted into an
approximate viewing angle \cite{SWW2000} and the torus opening angle can be
inferred from the distributions of quasars and radio galaxies.

\section{Summary}

The receding torus model provides a successful quantitative and physical
framework within which to unify broad- and narrow-line AGN. The
quantitative aspect is important for studies of the high-redshift Universe
since neither optically-selected QSOs nor radio/optically-selected radio
galaxies represent a fundamental class of object and their number densities
must be related to the total (radio-loud) AGN number density by accounting
for the missing population (narrow- and broad-line, respectively).

\end{document}